\documentclass[]{JHEP}
\usepackage{epsfig}
\def\DR{{\overline{\mbox{\tiny DR}}}}
\def\MS{{\overline{\mbox{\tiny MS}}}}

\preprint{DAMTP-2000-23 \\ Cavendish-HEP-00/02 \\ CERN-TH/2000-144}

\title{Naturalness Reach of the Large Hadron Collider in Minimal Supergravity}

\author{B.C.~Allanach$^*$, J.P.J.~Hetherington$^\dag$,
M.A.~Parker$^\dag$ and B.R.~Webber$^{\dag,\ddag}$\\
$^*$DAMTP, University of Cambridge, Wilberforce Rd, Cambridge CB3 0WA, UK\\
$^\dag$Cavendish Laboratory, University of Cambridge, Madingley Road, Cambridge, CB3 0HE, UK\\
$^\ddag$Theory Division, CERN, 1211 Geneva 23, Switzerland}


\abstract{We re-analyse the prospects of discovering supersymmetry at the LHC, 
in order
to re-express coverage in terms of a fine-tuning parameter and to extend the
analysis to scalar masses $(m_0)$ above 2~TeV. We use minimal supergravity (mSUGRA) 
unification
assumptions for the SUSY breaking parameters.
The discovery reach at high $m_0$ is of renewed interest because this region
has recently been found to have a focus point, leading to relatively low
fine-tuning, and because it remains uncertain how much of the region can be
ruled out due to lack of radiative electroweak symmetry breaking.

The best fine tuning reach is found in a mono-leptonic channel,
 where for $\mu>0, A_0=0$ and $\tan \beta=10$
 (within the focus point region), and a top mass of 174~GeV,
all points in mSUGRA with $m_0 < 4000$~GeV, with a fine
tuning measure up to 210 (500) are covered by the search, where the definition
of fine-tuning excludes (includes) the contribution from the top Yukawa
coupling. Even for arbitrarily high $m_0$, mSUGRA can be discovered through
 gaugino events, provided the gaugino mass parameter $M_{1/2} < 460 $~GeV. In
 this region, the mono-leptonic channel still provides the best reach.
}

\begin{document}

\section{Introduction}

A possibility for new physics beyond the standard model is supersymmetry
(SUSY). If fermionic generators are added to the bosonic generators of the
Lorentz group, the new space-time symmetry is supersymmetry.
As a result of exact
supersymmetry, all particles have a partner of equal mass but opposite
spin-statistics. Cancellations between bosonic and fermionic loops prevent
radiative corrections from driving scalar masses up to the highest scale
present,
assumed to be the GUT or Planck scale, $10^{16}$ to $ 10^{19} $~GeV,
solving the naturalness problem of the standard model. In addition, the
renormalised electromagnetic, weak, and strong couplings can be made to 
converge to an approximately common
value at the grand unification scale.

Since supersymmetry is not observed 
amongst the already discovered particles, it must be a broken
symmetry.
 The scale at which supersymmetry is
broken, $M_{\mbox{\scriptsize SUSY}}$, would be the typical mass of the as yet
undiscovered
superpartners of the standard model particles, and represents the scale at
which this new physics becomes relevant. Considerations of general
new physics beyond the standard model~\cite{newphys} can result
 in upper bounds on new physics scales
of order a few TeV. However, in supersymmetric models~\cite{HABER},
 $M_{\mbox{\scriptsize SUSY}}$ is expected to be 
at most around 1 TeV. If the SUSY breaking scale is
too large, then, unless there are large cancellations between SUSY breaking
parameters, electroweak symmetry breaking also will be of order
$M_{\mbox{\scriptsize SUSY}}$,
and the $W$ and $Z$ bosons would have masses inconsistent with their
measured values.


If $M_{\mbox{\scriptsize SUSY}}$ is at or below the TeV scale,
 then supersymmetric particles will almost
certainly be discovered at the Large Hadron Collider (LHC), being
built at CERN\@. Indeed, detailed studies of how the SUSY
parameters would be measured by the LHC general-purpose
 experiments ATLAS~\cite{ATLASTDR2} and
 CMS~\cite{CMSTP} have been made~\cite{baernew}. 

The simplest possible 
SUSY extension of the standard model, with a superpartner for each standard
model particle, and the addition of a second Higgs scalar doublet, is called
the minimal supersymmetric standard model, or MSSM\@. The most studied
sub-category of the MSSM is minimal supergravity,
mSUGRA\@. Supergravity, where supersymmetry is a local,
rather than a global symmetry, 
at one time motivated unification assumptions amongst the MSSM SUSY breaking
parameters,
reducing the number of parameters, from the hundred or so of
the MSSM, to just four, plus one sign. Currently, the suppression of flavour
changing neutral currents, not supergravity, is the main motivator
for these assumptions. The theory
is fully specified by these parameters together with those
 of the standard model.

As mentioned above, it
 is possible to avoid
the problem of large electroweak boson masses if there are
extra cancellations amongst the SUSY masses. However, such situations, where
the parameters of a theory
 are carefully tuned to avoid unphysical results, are often
thought to be unsatisfactory. Fundamental parameters, it is argued,
 should be independent, uncorrelated inputs.

These `naturalness' arguments are often quantified in terms of `fine
tuning'~\cite{FTref}. There are a few different fine-tuning measures
 \cite{measures} , and all
are intended to be measures of the degree of cancellation required between
fundamental parameters. A value of fine-tuning above which a theory
becomes unacceptable is often advanced, and used to support the
argument that $M_{\mbox{\scriptsize SUSY}}$ should be small.

The authors of \cite{anderson} discuss a more sophisticated measure of
fine-tuning and use it to assess the status of the MSSM if superparticles are
not found at the LHC\@. In \cite{casas} fine-tuning motivated upper bounds
on MSSM masses are
obtained using a complete one-loop effective potential. \cite{mar} discusses
the use of fine-tuning to compare different
high-energy supergravity scenarios, while \cite{wright} uses fine-tuning to
compare non-minimal supersymmetric models. 

Since its status as a possible 
solution of the naturalness problem of the standard model is one of
the main reasons for investigating supersymmetry, naturalness arguments
have increased relevance to studies of supersymmetry.

We therefore
contend that fine-tuning is a relevant way to compare SUSY models,
experiments, and
search channels, and is a useful measure of experimental discovery reach.
As experiments push the lower bounds on SUSY
parameters upwards, the minimum fine-tuning which SUSY can have
increases, and our confidence that the universe is supersymmetric at low
energies falls.
However, we do not believe that a high fine-tuning in itself can
be used to rule a theory out.

Within mSUGRA, the fundamental SUSY breaking parameters are boundary conditions on the running
SUSY breaking masses and couplings imposed at a high scale, usually taken to be
$10^{16}$~GeV. Physical masses of superparticles are obtained by evolving the
MSSM parameters to the weak scale using the renormalisation
group equations (RGEs). The RGE evolution of
minimal supergravity shows a `focus point' behaviour~\cite{focuspoints, focuspoints2}. A relatively
large region of
GUT scale parameters exists for which the RGE trajectories converge towards a small
range of measurable properties. Specifically, the renormalisation 
group trajectories of the mass squared of a Higgs
doublet ($m_{H_2}^2$) cross close to the electroweak scale. 

As a result, the electroweak symmetry breaking is insensitive to the GUT
scale SUSY breaking parameters~\cite{focuspoints}, and fine-tuning is smaller
than expected. This focus point corresponds to a region in which
the scalar SUSY breaking masses, governed by the mSUGRA parameter $m_0$,
may be large. 

Previous predictions of the discovery reach
of the LHC in mSUGRA parameter
space, using ISAJET~\cite{ISASUSY}, went only as far as $m_0 <
2$~TeV~\cite{ATLASTDR2}. The purpose of the present
investigation is to extend this reach to higher $m_0$
and to present it in terms of a
naturalness measure. We seek to determine
how the standard SUSY search channels perform in this region, where charginos
and neutralinos would be the dominant SUSY particles. For large $m_0$,
squarks, sleptons, and the heavy Higgs particle could avoid detection at the
LHC, and the determination of the $m_0$ reach of searches for these particles
would be an interesting further study.


In section~\ref{ft}, we discuss our fine-tuning measure, and which parameters to
include in its definition. We discuss in section~\ref{constr} the matter of the
electroweak symmetry breaking excluded region.
In section~\ref{MCS} we discuss SUSY search channels considered for use
at the LHC, and our simulation of the discovery reach using the
HERWIG~\cite{HERWIG,HERWIG2} event generator.
We present in section~\ref{results} the fine-tuning reach in each
channel, and the overall fine-tuning reach of the LHC.

\section{Fine tuning}
\label{ft}
At tree-level, in the MSSM, the $Z$ boson mass is determined to be~\cite{HABER}:

\begin{equation}
\frac{1}{2} M_Z^2 = \frac{m_{H_1}^2 - m_{H_2}^2 \tan^2 \beta}{\tan^2 \beta -
1} - \mu^2 \label{FTtree}
\end{equation}
by minimising the Higgs potential. $\tan \beta$ is the ratio of Higgs
vacuum 
expectation values (VEVs) $v_1/v_2$ and $\mu$ is the Higgs mass parameter in
the MSSM superpotential. 
In mSUGRA, $m_{H_2}$ has the same origin as the super-partner masses ($m_0$). Thus as
search limits put lower bounds upon super-partners' masses, the lower bound
upon $m_0$ rises, and consequently so does $|m_{H_2}|$. A cancellation is then
required between the terms of equation~\ref{FTtree} in order
to provide the measured value of $M_Z \ll |m_{H_2}|$. Various measures have
been proposed in order to quantify this cancellation~\cite{measures}.

The definition of naturalness $c_a$ of a `fundamental' parameter $a$ employed
in reference~\cite{focuspoints} is
\begin{equation}
c_a \equiv \left| \frac{\partial \ln M_Z^2}{\partial \ln a} \right|.
\end{equation}
From a choice of a set of fundamental parameters $\{ a_i \}$, the
fine-tuning of a particular model is defined to be $c=\mbox{max}(c_a)$.
Our initial choice of free, continuously valued, independent and fundamental mSUGRA
parameters also follows ref.~\cite{focuspoints}:
\begin{equation}
\{ a_i \} = \{ m_0, M_{1/2}, \mu(M_{\mbox{\tiny GUT}}), A_0, B(M_{\mbox{\tiny GUT}}) \}
\label{sugparm}
\end{equation}
where $M_{\mbox{\tiny GUT}} \sim 10^{16}$ GeV is the GUT scale. It is this selection which
gives rise to low fine-tuning for large $m_0$.

We have calculated $c$ numerically to one-loop accuracy in soft
masses, with two-loop accuracy in supersymmetric parameters.
Dominant one-loop top/stop corrections were added to
the Higgs potential and used to correct eq.~(\ref{FTtree}). The Higgs
potential was minimised at $Q= \sqrt{(m_{{\tilde t}_1} m_{{\tilde t}_2})}$,
where its scale dependence is small.

Full one-loop sparticle and QCD corrections were used to determine
$m_t(m_t)^\DR$~\cite{PBMZ} and the running of $\tan\beta$ was taken
into account~\cite{HabAndEsp} in order to calculate the Yukawa
couplings from fermion running masses. 
Fermion running masses were
determined at $M_Z$ by evolving them with 3-loop QCD $\otimes$ 1-loop QED.
$m_b(M_Z)^\DR$ and $m_\tau(M_Z)^\DR$ were determined by
including one-loop SUSY QCD and third family corrections~\cite{PBMZ}.
The $\overline{\mbox{DR}}$ Higgs vacuum expectation
value $v\equiv \sqrt{v_1^2 + v_2^2}$ was determined by the approximate
formula~\cite{PBMZ}
\begin{equation}
v(Q) = 248.6 + 0.9 \ln \left( \frac{m_{\tilde u_L}}{Q} \right)
\;\;\mbox{GeV}.
\end{equation}
One-loop top, gluino and squark corrections were
used~\cite{PBMZ} in order to deduce 
$\alpha_S^\DR(M_Z)$ from 
$\alpha_S^\MS(M_Z)=0.119$. 
We will examine the implications of relaxing the accuracy of the calculation
below. 

Note that our code does not yet
include 2-loop soft terms in the RGEs, 
finite corrections to the electroweak gauge couplings, a full
one-loop calculation of the Higgs vacuum expectation value,
or tadpole contributions to the Higgs potential. It is well known~\cite{casas}
that improving the accuracy of the calculation of $c$ at this level can make
a significant difference to its numerical value. Since its exact
quantitative interpretation is obscure anyway, 
this fact is not in conflict with the proposed {\em comparison} of
naturalness reaches in different channels.

We also consider the case where
the top Yukawa coupling $h_t(M_{\mbox{\tiny GUT}})$ is added to the list
 of fundamental parameters in eq.(\ref{sugparm}).
 In refs.~\cite{focuspoints,focuspoints2,incht}, it is shown that the focus
point scenario with heavy scalars has a large fine-tuning if $h_t$ is
included. The inclusion of $h_t(M_{\mbox{\tiny GUT}})$ into the definition of fine
tuning 
thus increases the naturalness reach of the LHC, as our results in section~\ref{results} show.

\section{Radiative electroweak symmetry breaking}
\label{constr}

Most plots of search reach in the $m_0$,$M_{1/2}$ plane found in the
literature \cite{ATLASTDR2,focuspoints,hbaeretc} show a large excluded
 triangular region, for large $m_0$ and small $M_{1/2}$. In this region,
mSUGRA does not have the required properties for radiative electroweak
symmetry breaking (REWSB).  However, the REWSB constraint, unlike most
of the {\small SUSY} spectrum, is very sensitive to details of the
calculation and input parameters, particularly the value of the
running top mass $m_t(m_t)$, which must be calculated from the pole
top mass in the renormalisation scheme being used.  The precise
relationship between the running and pole masses depends on
the masses of the superparticles~\cite{HabAndEsp},
and hence on the point in the {\small SUSY} parameter
space being investigated.

In version 7.14 of ISASUGRA, which was used in~\cite{ATLASTDR2,hbaeretc}
to generate SUSY masses and mixings and to obtain the REWSB excluded region,
the region above $m_0\sim 2M_{1/2}+1$ TeV was excluded for an input pole top
mass of 174~GeV.  Here one-loop RGEs were used and minimisation of the scalar
potential was performed at scale $M_Z$~\cite{hbpriv}.

\FIGURE{
\hbox{\epsfysize=8cm
\epsffile{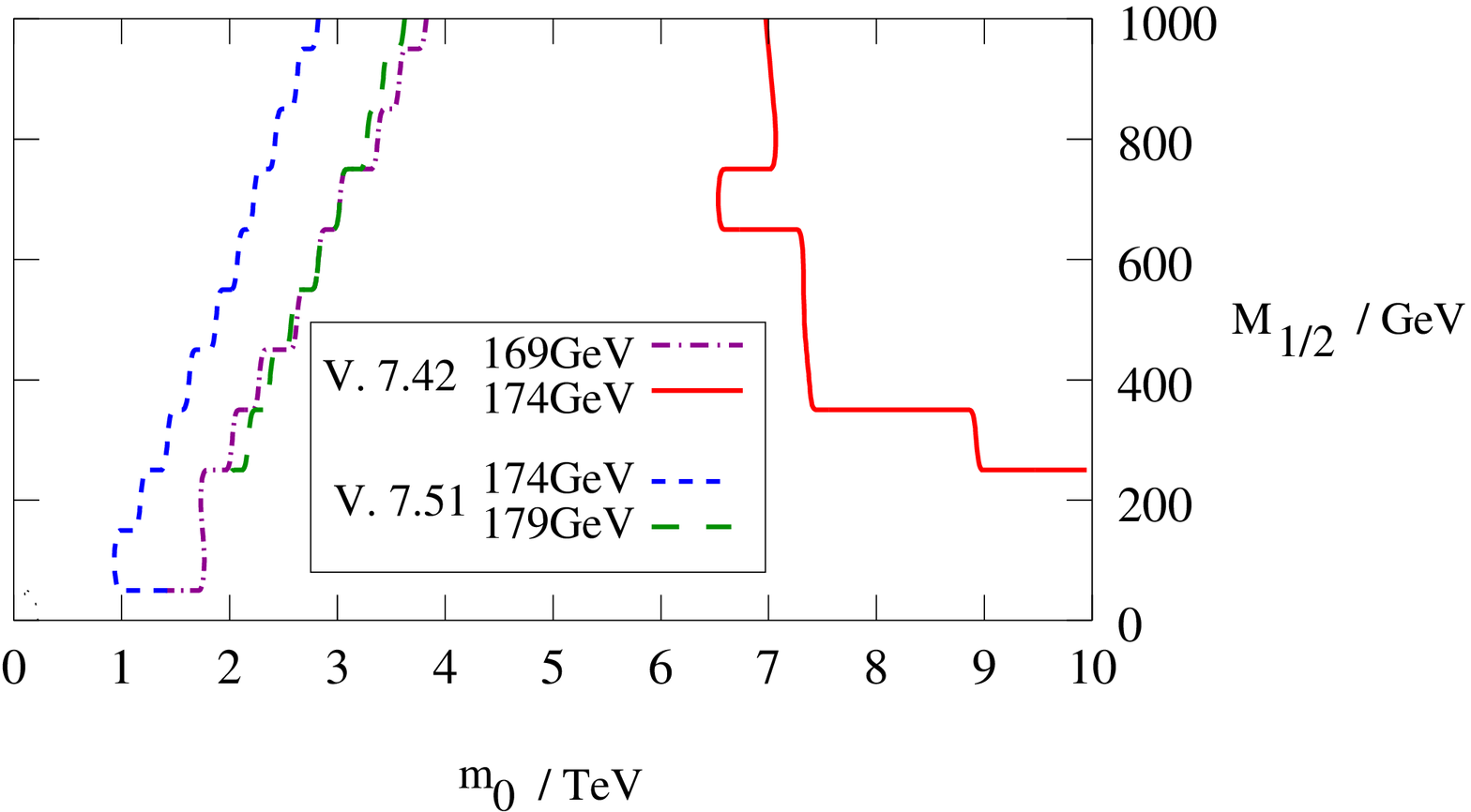}}
\caption{Dependence of the radiative electroweak symmetry breaking excluded
region on top mass, according to {\small ISASUGRA} versions 7.42 and 7.51,
for $\tan\beta = 10$, $\mu >0$ and $A_0 = 0$.
}
\label{EWSBISA}
}

Between 7.14 and more recent versions~\cite{ISASUSY}
such as 7.42, the excluded region shifted to very high $m_0$,
around 6~TeV for an input pole top
mass of 174~GeV, as shown in figure~\ref{EWSBISA}.
The REWSB constraint therefore no longer
provides a useful limit on the values of $m_0$ which should be
investigated. Here and in later versions,
the scalar potential
minimisation takes place at $Q= \sqrt{m_{{\tilde t}_L} m_{{\tilde
t}_R}}$~\cite{hbpriv}. 
In the most recent version of ISASUGRA, 7.51, the region has returned once
again to low values of $m_0$, as shown also in in figure~\ref{EWSBISA}. Here, 
two-loop corrections to the RGEs and one-loop sparticle mass corrections to
the inputs have been added~\cite{hbpriv}.

The REWSB exclusion region was also found using our fine-tuning calculation code,
where the physics included was as described in the previous section.
The dependence of the exclusion region on various aspects of the physics
is shown in figure~\ref{variety}.  We note again a strong dependence on
the top mass, and on the treatment of the running mass and strong coupling.

\FIGURE{
\hbox{\epsfysize=8cm
\epsffile{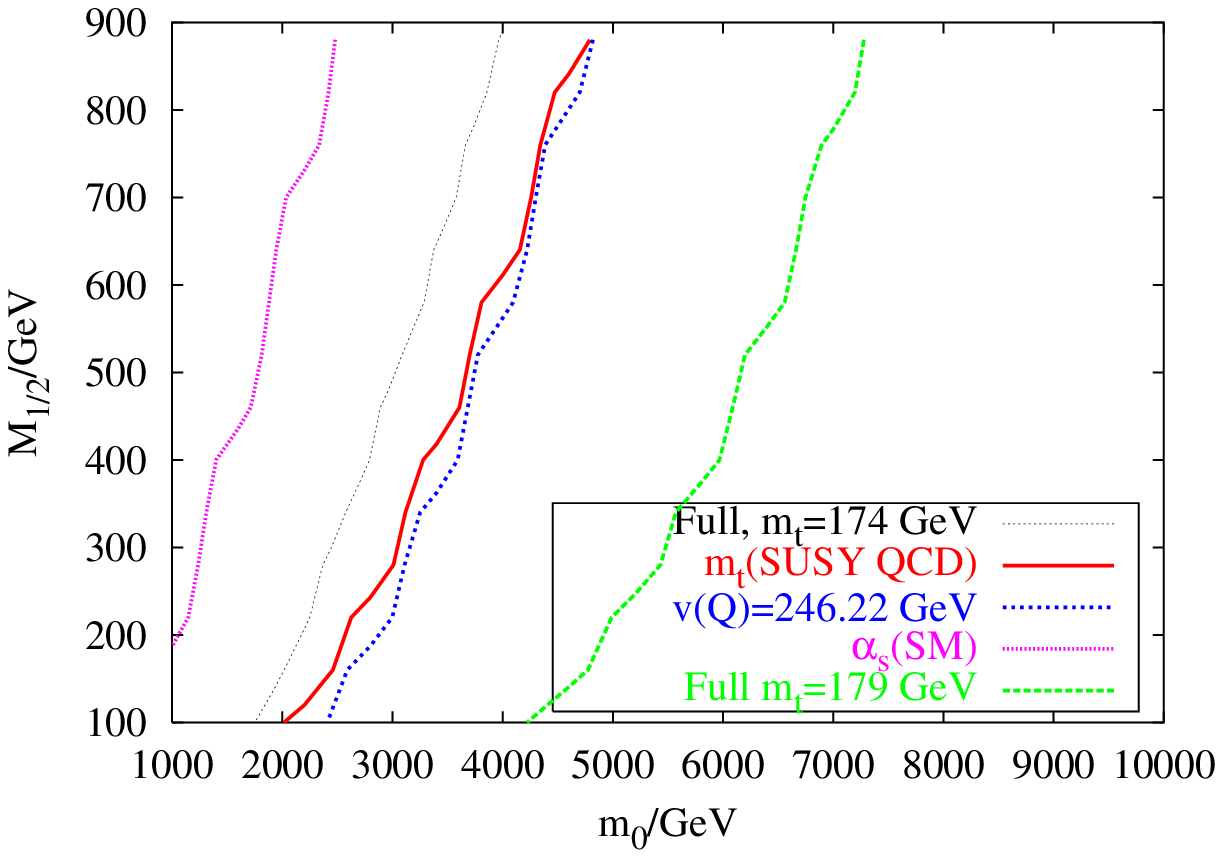}}
\caption{The dependence of the REWSB excluded region in our fine-tuning code
on various approximations for $m_t=174$ GeV, except for the curve marked
$m_t=179$ GeV. 
The curves shown are: the full calculations (`$m_t=174$ GeV', `$m_t=179$ GeV')
and calculations leaving out chargino and neutralino corrections to
the running top mass (`$m_t$(SUSY QCD)'), using only a constant value of the
Higgs VEV (`v(Q)=246.22 GeV'), and using 
only standard model corrections to the $\overline{\mbox{DR}}$
 value of $\alpha_S(M_Z)$ (`$\alpha_s$(SM)').
Regions to the right of the curves are ruled out by the REWSB
constraint. 
}
\label{variety}
}

Since the REWSB constraint is so heavily dependent upon the precise values
of input parameters and details of the calculation, there is no reason to
suppose that the exclusion region will remain stable under further refinements.
Hence we believe it should not be taken as a limit upon the
regions of mSUGRA parameter space to be investigated experimentally. Both our
own code and ISASUGRA could be subject to large corrections. 
We therefore use ISASUGRA7.42 to compute the discovery reach, since the
allowed region then extends up to values of $m_0=10000$ GeV (see
figure~\ref{EWSBISA}). As we shall show, the discovery contour in the
allowed region is not so sensitive to the version number or top mass.
To produce the REWSB excluded regions displayed in the plots of our results,
we used our fine-tuning code with the pole top masses specified in the captions.

\section{Monte-Carlo simulation of LHC discovery reach}
\label{MCS}

We now turn to the discussion of the LHC mSUGRA search.

In this study, as in \cite{hbaeretc}, the results of which were used in \cite{ATLASTDR2}, the SUSY discovery reach in the $m_0$, $M_{1/2}$ plane
was found through a variety of signals involving hard isolated leptons, jets
and
 missing transverse momentum. These cuts can be applied with a range
of cut energies $E_{cut}$, and are listed below. The cuts represent typical
SUSY discovery cuts that might be used at a general-purpose
LHC experiment, such as ATLAS~\cite{ATLASTDR2} or CMS~\cite{CMSTP}.

\begin{itemize}
\item
Missing transverse momentum ${/\!\!\!\!p_T}>E_{cut}$
\item
Either: At least 2 jets, with pseudo-rapidity $|\eta|<3.2$,
 and $p_T>E_{cut}$, using a
cone algorithm with cone-size 0.7 units of $\eta, \phi$.
Or: No jets with $p_T >$ 25~GeV, using the same jet-finding algorithm.
\item

Any final state ($\mu$ or $e$) leptons, with $|\eta|<2.0$ and $p_T> 20$~GeV,
and lying further in $\eta, \phi$ space than 0.4 units from the centre
of any 15~GeV jet with cone-size 0.4, and with less than 5 GeV of energy within 0.3
units of the lepton.

Channels with no leptons, one lepton, two leptons of opposite or the same sign, and three
leptons were investigated.

\item

In the one-lepton channel, an extra cut was imposed to reduce the background
due to standard model $W$ decay, involving the transverse mass:
\begin{equation}
M_T =  \sqrt{2 (| {\bf p_l} | |{\bf /\!\!\!\!p_T}| - {\bf p_l} \cdot {\bf
/\!\!\!\!p_T} ) } >
100\mbox{~GeV},
\end{equation}
where ${\bf p_l}$, ${\bf /\!\!\!\!p_T}$ are transverse two-component momenta,
for the lepton and
missing $p_T$ respectively.

\item

\begin{equation}
S_T = \frac{2 \lambda_2}{\lambda_1 + \lambda_2} > 0.2,
\end{equation}
where $\lambda_i$ are the eigenvalues of the matrix
$S_{ij} = \Sigma_{ij} p_i p_j $,
the sum being taken over all detectable final state particles, and
$p_i$ being the two-component transverse momentum of the particle. $S_T$ is often called the transverse circularity or transverse sphericity.
\end{itemize}

The search channels used are shown in table~\ref{channels}.

\TABULAR{|c|c|c|}{
\hline
Jets & Leptons & Label \\
\hline
$>$ 1 & 0  & j0 \\
& 1 & j1 \\
& 2 (Opposite sign) & jos \\
& 2 (Same sign) & jss \\
& 3 & j3 \\
\hline
0 & 2 & v2 \\
& 3 & v3 \\
\hline
}
{Channels in which the SUSY discovery 
reach has been investigated. \label{channels}
}

The mSUGRA events were simulated by employing the {\small ISASUSY} part of the
{\small ISAJET7.42}
package~\cite{ISASUSY} to calculate sparticle masses and branching ratios, and
{\small HERWIG6.1}~\cite{HERWIG, HERWIG2} to simulate the events themselves.
The expected SUSY signal was generated, together with the backgrounds due to
standard model top anti-top and $W$ plus jet events. The $W$ background was
generated using events with one of the jets produced in the hard subprocess,
and the rest in QCD parton cascades. This produces an underestimate
of the $W$ background, as discussed in \cite{bryansw}, and the results were
rescaled accordingly.

The discovery limit is set
at
values of $m_0$ and $M_{1/2}$ where $S/\sqrt{B}>5$ and $S>10$ where S and B
are the expected number of events in the
the SUSY signal and total background. The former constraint is an
approximation to the requirement that the total number of observed events
in some experiment will be significantly above the
background at the 5 $\sigma$ level.

For each value of $E_{cut}$,
the minimum final state transverse momentum in
the hard subprocess was selected to obtain some events passing the cuts, within realistic computer
time constraints. Where the
minimum transverse momentum had to be increased above $E_{cut}$, 
results from the j0 channel were
used to obtain linear correction factors to
compensate for the resulting underestimate
 of the background.

In order to find the hard leptons, and to determine  $\bf/\!\!\!\!p_T$
and $S_T$, the Monte
Carlo output was examined directly, particles with $|\eta|>5.0$ being
ignored.
For the jet count and lepton isolation
check, a calorimeter simulation was used, with hadronic resolution of 
 $ 70\% / \sqrt{E}$,
electromagnetic resolution of $10\%/\sqrt{E}$ and extending to
pseudo-rapidities up to 5.0. 
These parameters have been selected to simulate a general-purpose
experiment at the LHC.

Calculations were made for mSUGRA with $A_0=0$, $\tan \beta=10$,
pole top mass 174
GeV, $\mu>0$ and ${\mathcal L}=10$
fb$^{-1}$ of integrated luminosity.
The integrated luminosity chosen is equivalent to one year of running in the low
luminosity
mode, $10^{33}$ cm$^{-2}$ s$^{-1}$.
We note that the LHC experiments expect to collect around
300 fb$^{-1}$ of data.

A low-statistics calculation of the discovery reach with a top mass of 179~GeV
was also made, to determine whether the reach is independent of the top
mass. There is a small decrease in the production cross section for $t\bar t$
background with increasing top mass, and a compensating increase in the energy
of the resulting leptons and jets, which causes a greater proportion to pass
the selection cuts. Thus for this study the background could safely be assumed to
be independent of the top mass. 

\section{Results}
\label{results}

Preliminary results on
the naturalness reach of the LHC for the j1 channel, with $E_{cut} =
400$~GeV, were presented in~\cite{ourproceedings}. The top
Yukawa coupling was not included in the definition of fine-tuning. Here we
obtain the discovery limit in all the channels in table~\ref{channels}, for
several values of $E_{cut}$, and with both definitions of fine-tuning
(with/without the top Yukawa coupling).

Table~\ref{backgtable} shows the background for each generated channel, the total
background,
and the resulting expected number of supersymmetry events required to fulfill
the discovery criteria detailed above. No discernible background in the jet veto channels (v2,v3) was
obtained.

\TABULAR{|c|c|c|c|}{
\hline
Channel & $E_{cut}$ / GeV & Total no of &
Required no of \\
& & background events & SUSY Signal events \\
\hline
j0 & 100 & $ 5.3 \times 10^5 $ & $ 3.6 \times 10^3 $ \\
j0 & 200 & $5.2 \times 10^4 $ & $1.4 \times 10^3 $ \\
j0 & 300 & $ 7.4 \times  10^3$ & 431 \\
j0 & 400 & $ 1.6 \times 10^3$ & 201 \\
j0 & 500 & 451 & 107 \\
\hline
j1 & 100 & $ 1.3 \times 10^4 $ & 579 \\
j1 & 200 & 394 & 102 \\
j1 & 300 & 35 *& 29 \\
j1 & 400 & 4 & 10 \\
j1 & 500 & 1 & 10 \\
\hline
j3 & 100 & 36* & 30 \\
j3 & 200 & 20* & 22 \\
j3 & 300 & $<1$* & 10 \\
j3 & 400 & $<1$ & 10 \\
j3 & 500 & $<1$ & 10 \\
\hline
jss & 100 & 115* & 54 \\
jss & 200 & 7*  & 13 \\
jss & 300 & 4* & 10 \\ 
jss & 400 & 4 & 10 \\
jss & 500 & 2 & 10 \\
\hline
jos & 100 & $ 7.4 \times 10^3$ * & 431 \\
jos & 200 & 173 * & 65 \\
jos & 300 & 12 * & 17 \\
jos & 400 & 5 & 11 \\
jos & 500 & 2 & 10 \\
\hline
}
{
Total background for SUSY discovery channels. Also
shown is the expected number of SUSY events required for discovery. 
Backgrounds marked
with an asterisk $(*)$ have been corrected for the use of a minimum
subprocess $p_T$ greater than $E_{cut}$, as described in section 4.
\label{backgtable}
}

\TABULAR{|c||c|c||c|c||c|c||c|c||c|c|}{
\hline
Cut Type & $E_{cut}$ & $c$ & $E_{cut}$ & $c'$ & $E_{cut}$ &
$c_{h_t}$ & $E_{cut} $
 & $c'_{h_t}$ 
& $E_{cut}$ & $M_{1/2}$ limit \\
\hline
j0 & 300 & 120 & 300 & 230 & 400 & 375 & 400 & 375 & 200 & 350 \\
j1 & 400 & 210 & 400 & 260 & 400 & 500 & 400 & 550 & 300 & 460 \\
jos & 300 & 85 & 300 & 190  & 300 & 240 & 300 & 270 & 200 & 300 \\
jss & 200 & 110 & 200 & 215 & 200 & 240 & 200 & 270 & 200 & 350 \\
j3 & 100 & 70 & 100 & 130 & 100 & 150 & 100 & 155 & 100 & 280  \\
\hline
}
{
Values of the cut on missing momentum and jet energy which give the best
fine-tuning limits $c$ and $c_{ht}$, and the $M_{1/2}$ reach for large
$m_0 >$ 4000~GeV, for each of the supersymmetry discovery channels.
$c$ is the fine-tuning limit as defined in the text, and $c_{h_t}$
is the same with the fine-tuning with respect to the top
Yukawa coupling included, using a top mass of 174~GeV to calculate the REWSB
excluded region and fine-tuning. $c'$ refers to the same quantities for a top
mass of 179~GeV. $E_{cut}$ and $M_{1/2}$ are given in GeV.
\label{resultstable}
}

For each of the channels, and each value of $E_{cut}$, a plot has been used
 to determine the
fine-tuning reach with both definitions of fine-tuning. The fine-tuning reach
 is the largest fine tuning where the fine-tuning contour is completely
 within the discovery limit, for $m_0 < 4000$~GeV.
The values of $E_{cut}$ giving the best fine-tuning reach,
and the corresponding reach, are
shown in table~\ref{resultstable}. The results using $m_t=174$~GeV and
 $m_t=179$~GeV to calculate the fine-tuning and REWSB exclusion are given.
 $E_{cut} = $100, 200, 300, 400 and 500~GeV
 were investigated. No results can be presented for the jet veto channels, 
as the number of events in these channels is too small. The
 reach in these channels could be obtained by considering specific SUSY
 processes, but is expected to be very limited.

\FIGURE{
\hbox{\epsfysize=10cm
\epsffile{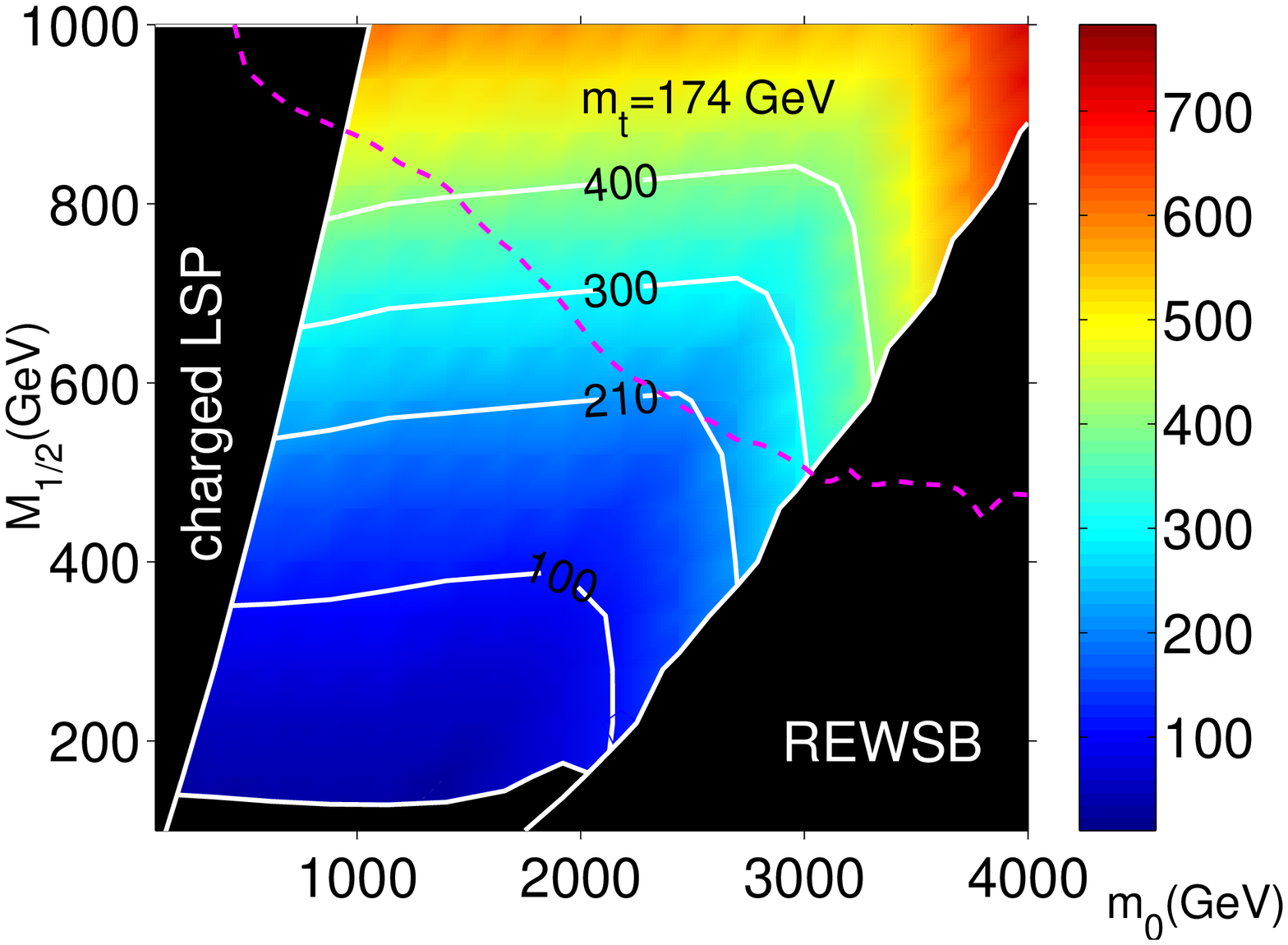}}
\caption{Naturalness reach at the LHC for $A_0=0$, $\tan \beta=10$,
$\mu>0$, $m_t = 174$~GeV in minimal SUGRA\@. The fine-tuning is represented by the background
density, as measured by the bar on the right. White contours of fine-tuning
are also presented. Fine
tuning with respect to the top Yukawa coupling is neglected. The dashed line is
the LHC expectation SUSY discovery contour in the j1 
channel described in the text for a luminosity of ${\mathcal L} =
10$~fb$^{-1}$, and with $E_{cut}=$~400~GeV. The excluded region, filled black,
along the
left hand side of the plot is due to the cosmological requirement that the
lightest supersymmetric particle be neutral, while that along the bottom of
the plot is due to the experimental lower bound on the chargino mass. The
excluded region on the right hand side of the plot is due to lack of
radiative electroweak symmetry breaking.
}
\label{fig:ftreachpos}
}

\FIGURE{
\hbox{\epsfysize=10cm
\epsffile{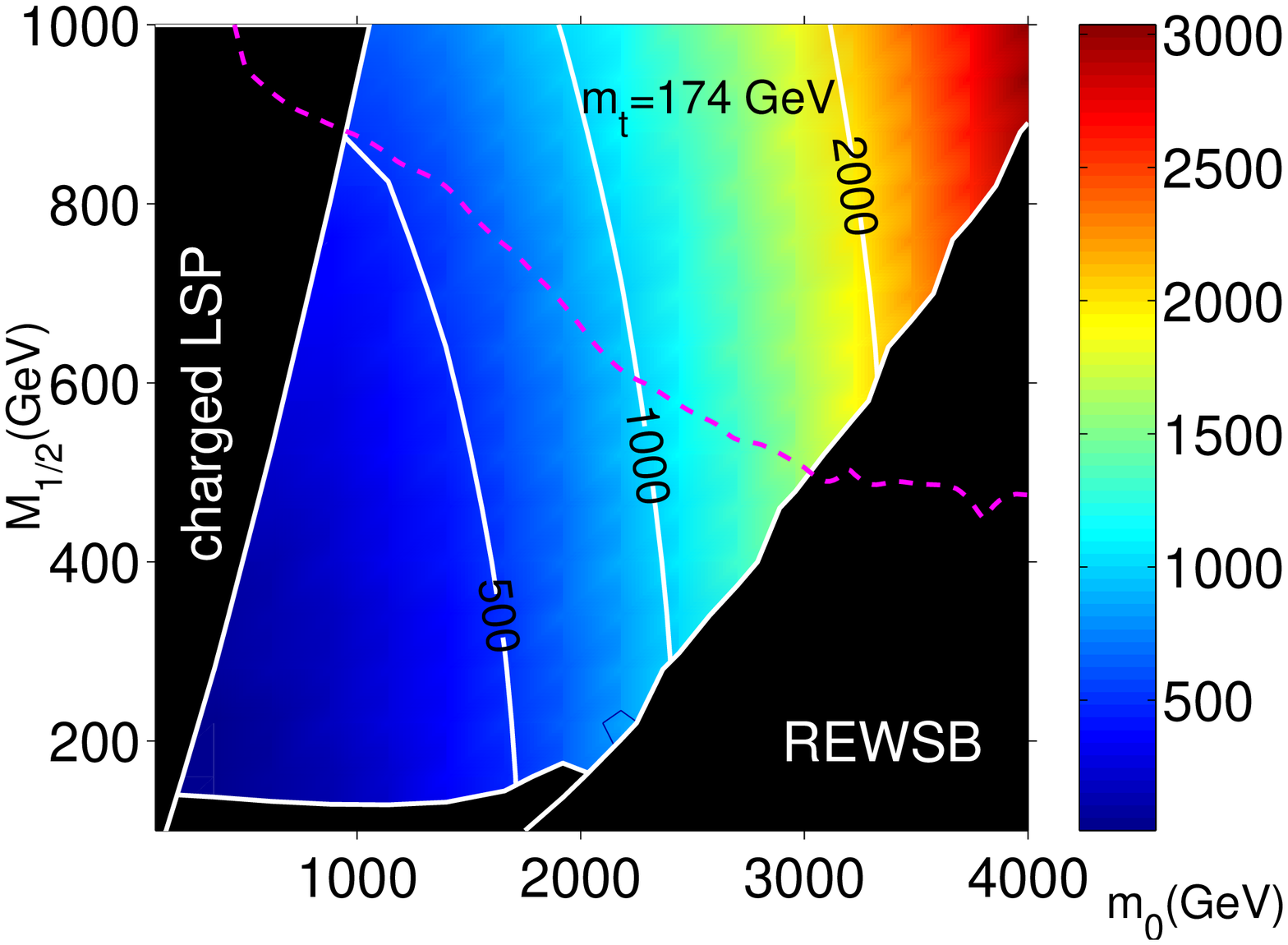}}
\caption{ As in figure~\protect\ref{fig:ftreachpos}, but with fine
tuning with respect to the top Yukawa coupling included.
}
\label{fig:ftreachyt}
}

Figures~\ref{fig:ftreachpos} and~\ref{fig:ftreachyt} show the best obtainable
discovery limit using the channel providing the best reach,
the $E_{cut}$ = 400~GeV j1 channel,
as a dashed
line in the $(m_0,M_{1/2})$ plane, and the naturalness density without and
with the top Yukawa coupling included, respectively. The resulting
fine-tuning limits, as defined above, are 210 and 500 units.
Excluded regions due to experimental limits on the chargino mass, and
due to the cosmological requirement that the LSP be neutral, are included in
the figures. The current limit on the light Higgs mass does not exclude any
of the region of mSUGRA parameter space illustrated, due to the large value
of $\tan \beta$ used here.

\FIGURE{
\hbox{
\epsfxsize=14cm
\epsffile{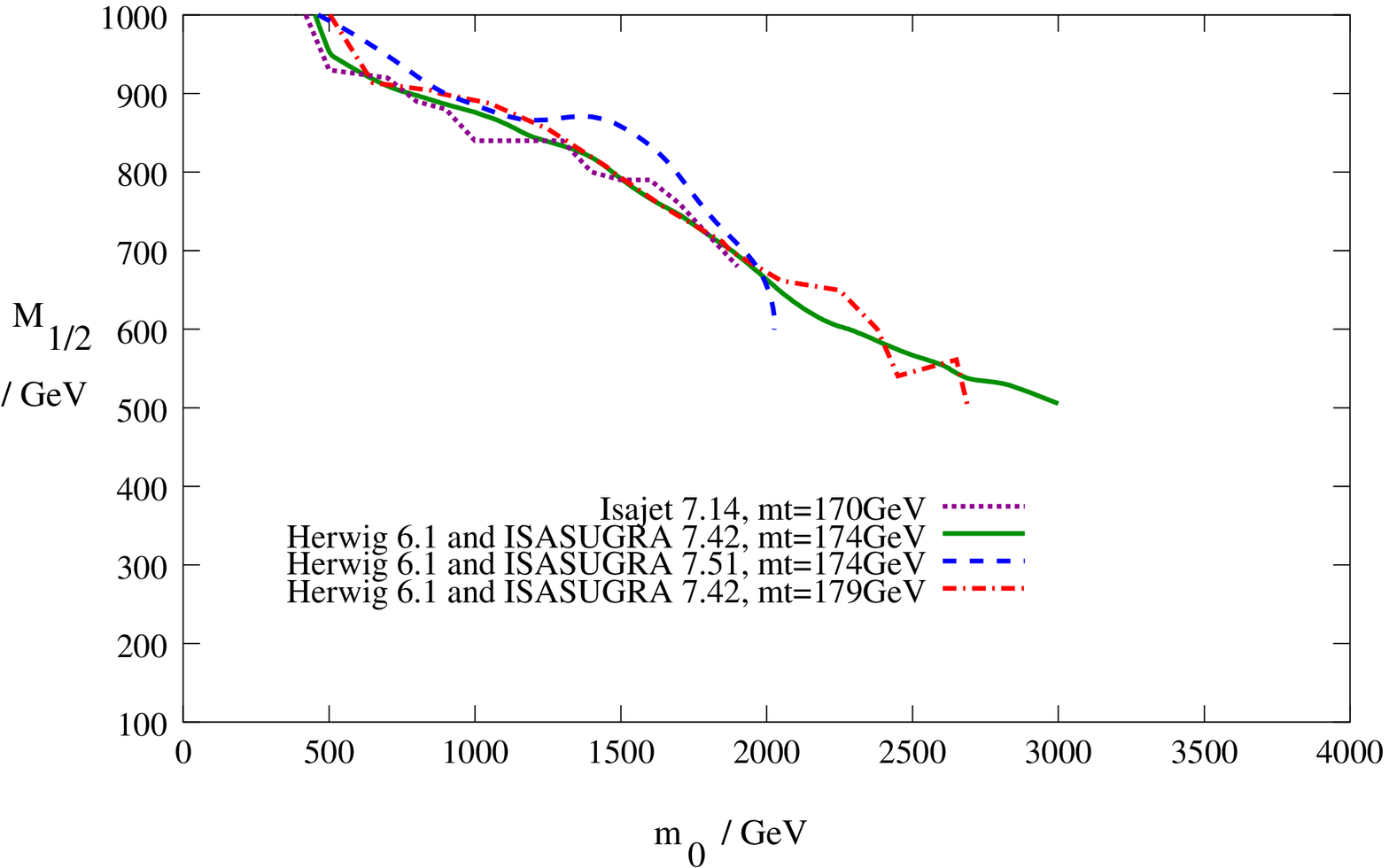}}
\caption{Reach at the LHC for $A_0=0$, $\tan \beta = 10$, $\mu>0$, 
in minimal supergravity. Solid: {\small HERWIG} results with $m_t =$174~GeV,
with {\small ISASUGRA} version 7.42 to generate the SUSY spectrum and decays.
Dot-dashed: The same with $m_t =$179~GeV.
Short-dashed: {\small ISAJET} results of \cite{ISASUSY} (with $m_t =$ 170~GeV).
Long-dashed:  {\small HERWIG} results with $m_t =$174~GeV, using
{\small ISASUGRA} version 7.51 to generate the SUSY spectrum and decays.
}
\label{fig:consistency}
}

Figure~\ref{fig:consistency}  shows
consistency of our results for $m_0<2$~TeV 
in the j1 channel with those obtained using
{\small ISAJET 7.14} in~\cite{hbaeretc}.
The same figure shows the discovery reach calculated with
{\small ISAJET7.51} (and lower
statistics). We see that
that the differences in approximation between the various
ISAJET versions 
which shifted the REWSB
exclusion, 
as discussed in section~\ref{constr},
do not appreciably alter the discovery reach.

Also shown in figure~\ref{fig:consistency} is
the discovery reach for a 1$\sigma$ increase in the top mass, $m_t=179$ GeV.
Unlike the REWSB exclusion region, the discovery contours are almost identical.
We may therefore use our high-statistics
$m_t=174$~GeV {\small ISAJET 7.42} discovery contour to calculate the
fine-tuning reach for $m_t=179$. As shown in table~\ref{resultstable} and
figure~\ref{mt179noyt}, the reach extends to a fine-tuning value of 260.

\FIGURE{
\hbox{\epsfysize=10cm
\epsffile{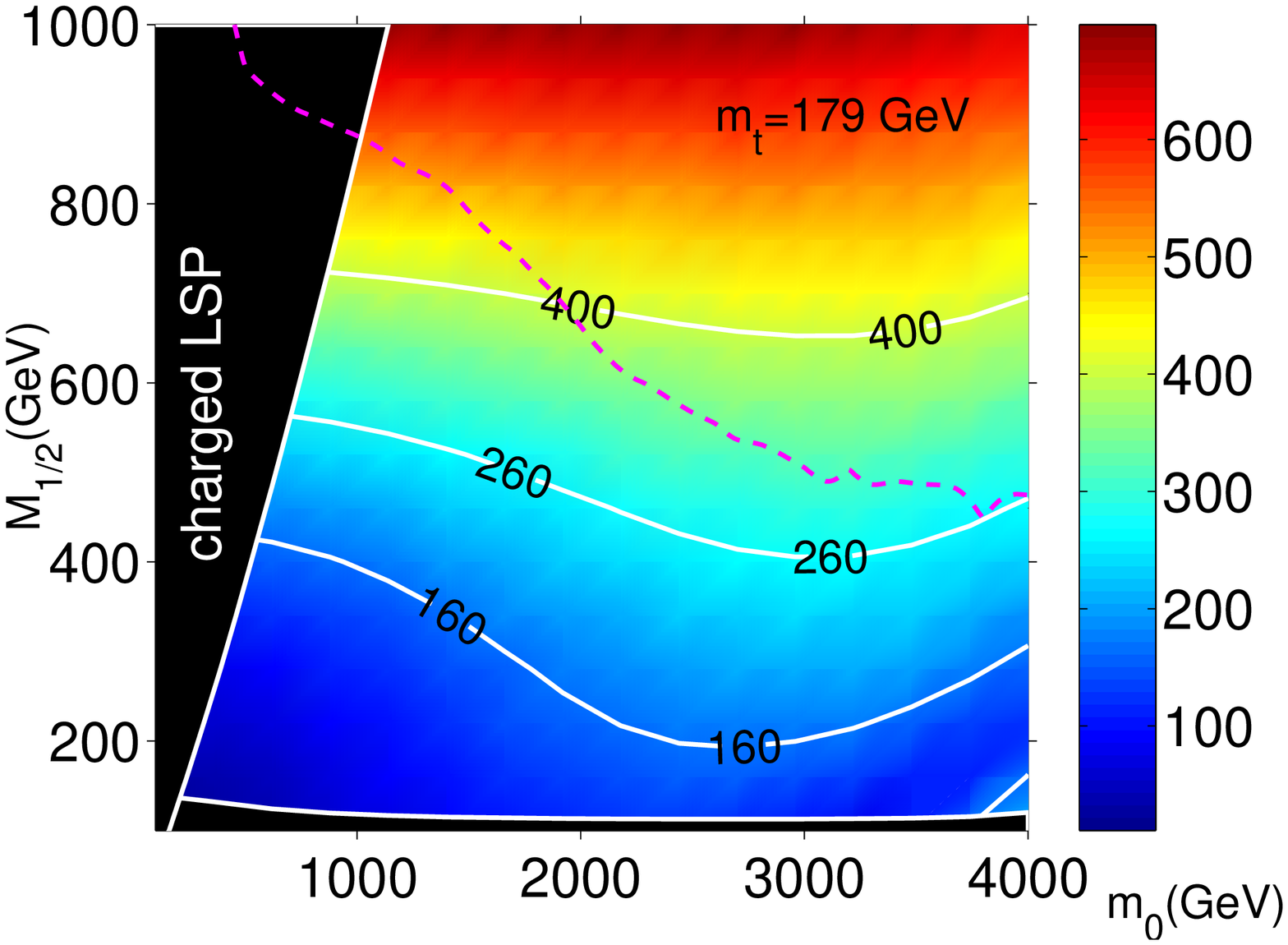}}
\caption{As in figure~\protect\ref{fig:ftreachpos}, but with a top mass of 179 GeV used to calculate
the fine-tuning and REWSB excluded region.}
\label{mt179noyt}
}

For $m_0 > 4000$~GeV
the discovery reach in $M_{1/2}$ becomes roughly independent
of $m_0$. Here, the SUSY processes involved are dominated by the gauginos,
the squark masses being very high (larger than 3 TeV), as illustrated in
figure~\ref{squark}. This figures also shows that scalar production,
as a fraction of total SUSY processes, decreases at high $m_0$, being typically
a percent around $m_0 \sim 3$ TeV. However, it is clear that at values of $M_{1/2}$
lower than the SUSY discovery contour but at $m_0 > 2$ TeV, it is possible to
produce scalar SUSY particles at the LHC. 
A discussion of how to 
attempt to obtain a limit on the region of mSUGRA parameter space
where the squarks in particular (rather than SUSY in general) can be
discovered is beyond the scope of this paper. 
\FIGURE{
\hbox{\epsfysize=8cm
\epsffile{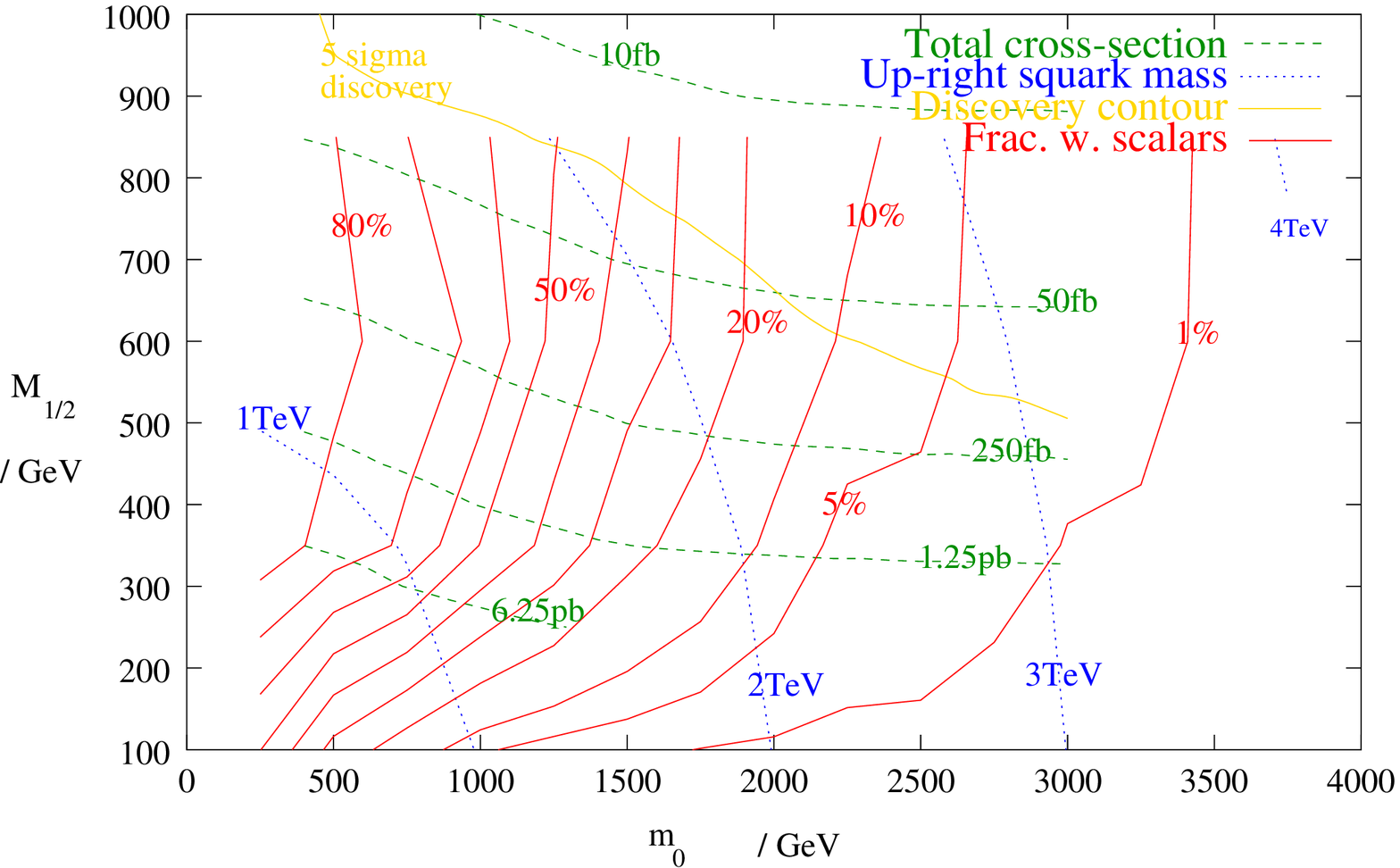}}
\caption{Scalar production in
mSUGRA. The contours represent the total SUSY cross-section,
the mass of a typical squark, (the up-right), the overall SUSY discovery
contour, and the fraction of SUSY processes involving a squark
or slepton.}
\label{squark}
}

We terminate the calculation of the fine-tuning and
search-reach at $m_0 =$~4000~GeV, since for higher values of $m_0$ the
$M_{1/2}$ discovery reach gives an adequate representation of the overall
SUSY discovery power, through the gauginos, in each channel. This expression
of the reach is also shown in table~\ref{resultstable}.

\section{Summary}
	
In this paper, we have obtained the discovery reach of the LHC into mSUGRA
parameter space, using the new supersymmetry routines in {\small HERWIG}. Where
our investigation repeats the calculation of~\cite{hbaeretc}, this provides a
useful
check on the consistency of the two Monte-Carlos. In addition, our use of the
latest software for calculating the mSUGRA spectrum updates the old results,
and allows us to move into the region of high scalar masses $m_0$.  It has
been suggested that a focus point gives this region increased naturalness,
and the extent to which radiative electroweak symmetry breaking excludes
high $m_0$ remains uncertain, so this region should be explored thoroughly.
We demonstrate
that even for arbitrarily high $m_0$, the standard SUSY searches at the 
LHC can discover
supersymmetry, through events involving gauginos, provided they
are not too heavy ($M_{1/2} < 460 $).

We have introduced the possibility of using
fine-tuning as a quantitative way to compare the discovery reach of various
channels. Fine-tuning can provide physicists with a quantitative measure of
discomfort with a theory, which increases as the  experimental bounds are
improved. Since it is this disquiet which leads, in the end, to the
abandonment of a theory such as supersymmetry, the fine-tuning reach
represents the potential of these discovery channels
for removing mSUGRA from the list of candidate theories for physics beyond
the standard model. The work could be repeated using some
different high-energy unification assumptions, instead of mSUGRA\@. The mSUGRA 
reach into $A_0$ and $\tan \beta$ could also be investigated.

The best fine tuning reach is found in a
mono-leptonic channel,
where for $\mu>0, A_0=0$,$\tan \beta=10$ 
 (within the focus
point region), and $m_t = 174$~GeV, 
all points in the $m_0$,$M_{1/2}$ plane with $m_0 < 4000$~GeV and a fine
tuning measure up to 210 (500) are covered by the search, where the definition
of fine-tuning excludes (includes) the contribution from the top Yukawa
coupling. For $m_0 > 4000$, all values of $M_{1/2} < 460$~GeV are covered by
the search. 

\acknowledgments
Part of this work was produced using the Cambridge University High Performance
Computing Facility. BCA would like to thank K.~Matchev for valuable discussions
on checks of the numerical results. JPJH would like to thank C.G.~Lester for useful
discussions and CERN Theory Division for hospitality.
The authors also thank G.G.~Ross for helping to motivate this study.

This work was funded by the U.K. Particle Physics and
Astronomy Research Council.

\end{document}